\documentstyle [12pt,epsfig]{article} 
\makeatletter
\@addtoreset{equation}{section}
\makeatother

\newcommand{\ba}{\begin{eqnarray}}
\newcommand{\ea}{\end{eqnarray}}

\begin{document}
\newcommand{\BS}{\bigskip}
\newcommand{\SECTION}[1]{\BS{\large\section{\bf #1}}}
\newcommand{\SUBSECTION}[1]{\BS{\large\subsection{\bf #1}}}
\newcommand{\SUBSUBSECTION}[1]{\BS{\large\subsubsection{\bf #1}}}

\begin{titlepage}
\begin{center}
\vspace*{2cm}
{\large \bf Forces between electric charges in motion: Rutherford scattering, circular Keplerian orbits,
 action-at-a-distance and Newton's third law in relativistic classical electrodynamics}
\vspace*{1.5cm}
\end{center}
\begin{center}
{\bf J.H.Field }
\end{center}
\begin{center}
{ 
D\'{e}partement de Physique Nucl\'{e}aire et Corpusculaire
 Universit\'{e} de Gen\`{e}ve . 24, quai Ernest-Ansermet
 CH-1211 Gen\`{e}ve 4.}
\end{center}
\begin{center}
{e-mail; john.field@cern.ch}
\end{center}
\vspace*{2cm}
\begin{abstract}
   Standard formulae of classical electromagnetism for the forces between electric charges in motion
  derived from retarded potentials are compared with those obtained from a recently
   developed relativistic classical electrodynamic theory with
   an instantaneous inter-charge force. Problems discussed include small angle 
  Rutherford scattering, Jackson's recent `torque paradox' and circular Keplerian orbits. 
    Results consistent with special relativity are obtained only with an instantaneous 
   interaction. The impossiblity of stable circular motion with retarded fields in either
   classical electromagnetism or Newtonian gravitation is demonstrated.

\end{abstract}
\vspace*{1cm}{\it Keywords};  Special Relativity, Classical Electrodynamics.
\newline
\vspace*{1cm}
\end{titlepage}

\SECTION{\bf{Introduction}}
   Classical electromagnetism (CEM), in particular Maxwell's equations, played an important
  role in the development of special relativity (SR). This does not however mean that CEM, 
  even as presented in modern text books, is a theory fully compatible with SR and
  and devoid of interpretational problems. This point was strongly emphasised by
  Feynman~\cite{Feyn1}. On the other hand, the most successful physical theory of the
  20th century, as much for its powers of prediction as for its precise experimental
  verification, is quantum electrodynamics (QED)~\cite{QED}
 . Again, as stressed by Feynman~\cite{Feyn2},
  the elements of this theory are very simple: charged particles and real or virtual 
  photons described by appropriate wavefunctions or propagators and a single coupling
  constant, the elementary electric charge. In QED the mechanical forces between charges
  are mediated by the exchange of space-like virtual photons. As discussed, in detail, in a 
  recent paper by the present author~\cite{JHF1}, the exchange of such photons is 
  instantaneous in the centre-of-mass (CM) frame of two interacting charges. This is a 
  consequence of the relativistic formula for velocity in terms of kinematical quantities:
   $ v = p c^2/E$. Energy-momentum conservation shows that E$(\gamma^*)$ vanishes whereas, in general,
   p$(\gamma^*)$ is non-zero for a virtual photon, $\gamma^*$, in the CM frame of any scattering 
   process, implying that its velocity
   is infinite. This conclusion is confirmed by the study, in ~\cite{JHF1}, of the invariant amplitude in momentum
    space for M\o ller scattering, $e^- e^- \rightarrow e^- e^-$, and its 
    Fourier transform.
Recently, convincing experimental evidence has been obtained~\cite{Kohletal,JHFnrf}
  
 for the non-retarded nature
   of `bound' magnetic fields with $r^{-2}$ dependence, (associated in QED with virtual photon exchange)
   in a modern version, probing small $r$ values, of the Hertz experiment~\cite{Hertz} in which the electromagnetic
   waves associated with the propagation of real photons (fields with $r^{-1}$ dependence) were
   originally discovered. 

    \par There is no reason to suppose that the space-like virtual photons responsible for the
    electromagnetic forces in  M\o ller scattering should not also be at the origin of the 
    electromagnetic
    forces between the electrons in two current-carrying conductors separated by macroscopic
   distances. Thus there should be a close connection between CEM and QED. Just this connection
   was explored in the second part of~\cite{JHF1}. In the first part of this paper all of the 
   mechanical equations of CEM, as well as Maxwell's equations, were derived from Coulomb's
   inverse-square force law, SR and Hamilton's Principle. The equations describing inter-charge
   forces, although consistent with standard CEM formulae in the limit where O($\beta^2$) and higher
   correction terms are neglected, differ significantly from them when such terms are included.
   The main aim of the present paper is a comparison of the RCED (standing for Relativistic
   Classical Electro-Dynamics) formulae from~\cite{JHF1} with the conventional text book formulae
   of CEM, to be found, for example, in~\cite{PP,LLCF,Jackson}. Indeed it is found that the
   latter, corresponding, for moving charges, to retarded potentials and fields do not respect
   the constraints of SR. As will be seen, this is one source of the previously unresolved
   problems of CEM which Feynman found to be of such great interest.
    \par The other source of problems and paradoxes in CEM is that electromagnetic forces 
   apparently do not respect Newton's third law. An explanation in terms of non-mechanical
  momentum and angular momentum residing in static electric and magnetic fields has been attempted
  by many authors~\cite{RHR,SJ,CVV,Furry,GGL,VH,ODJ} over the last half-century. The present
  author's opinion
  is that this is not the correct explanation, and that Newton's third law is actually
  always respected
  by electromagnetic forces, without the need to introduce non-mechanical degrees
  of freedom. One important point in this connection is the instantaneous nature of
  electromagnetic forces, a second is the correct treatment of relativistic effects, and the
  third is the understanding that the mass of an object is only a constant in the complete
  absence of interactions, which is evidently not the case for objects undergoing mutual
  electromagnetic interactions\footnote{The concept of an effective mass for a particle, depending
  upon the strength of the interactions it undergoes, is a familiar one in the theory of conduction
  in the solid state. The relation between the energy and momentum of a conduction electron
  can be markedly different from that for a free electron, for those occupying energy levels near to the upper
  or lower boundaries of a conduction band~\cite{CE}.} . 
  \par In fact it is clear that when all physical quantities are properly defined Newton's
  third law {\it must} be obeyed because it is a necessary consequence of Newton's first law,
  (the law of inertia) and the definition of `force' provided by the relativistic
  generalisation of Newton's second law. The proof of
   this is now briefly sketched. Given a dynamical theory
  it is always possible to obtain an equation such as:
  \begin{equation}
   \dot{\vec{p}}_O = \vec{f}_O(X,\alpha_i,\alpha_j...)
  \end{equation}
 where $\vec{p}_O$ is the relativistic momentum (see (1.6) below) of an object O, the dot
  denotes a time
 derivative and $\vec{f}_O$ is a known function of the space-time position $X \equiv (ct;\vec{x})$ 
  of O and some fixed parameters $\alpha_i,\alpha_j...$. The vector $\vec{f}_O$ is the relativistic
  generalisation of what is defined
 by Newton's second law as a `force'. Actually, however, this naming operation is the entire
  physical and logical content of the second law. The dynamics is completely defined by the
  knowledge of the time derivative of the relativistic momentum (1.1). Nothing is
  gained, except in an anthropomorphic sense, by baptising the right side of this equation a
  `force'. If it happens that $\vec{f}_O = 0$ in (1.1) then:
  \begin{equation}
 \vec{p}_O = {\rm constant}
 \end{equation}
  which is Newton's first law. Suppose now O is composed of two parts O$_1$ and O$_2$ so that:
   \begin{equation}
  \vec{p}_O =  \vec{p}_{O_1} +  \vec{p}_{O_2} =  {\rm constant}
 \end{equation}
 Differentiating (1.3) with respect to time and transposing gives:
  \begin{equation}
\dot{\vec{p}}_{O_1} = -\dot{\vec{p}}_{O_2}
 \end{equation}
 or, using (1.1)
  \begin{equation}
  \vec{f}_{O_1} = -  \vec{f}_{O_2}
 \end{equation}
 which is Newton's third law.  It is clear that it is a necessary consequence of the first law (1.2)
 and the definition (1.1) of `force' provided by the second law. The third law must then be obeyed if the
  relativistic momentum is correctly defined. 
  \par The general definition of the relativistic momentum of an object is:
  \begin{equation}
  \vec{p} \equiv m(X) \frac{d X}{d \tau} = m(X) \gamma \vec{v}
 \end{equation}
  where $\vec{v}$ is the object's velocity, $\tau$ its proper time, $\gamma \equiv 1/\sqrt{1-\beta^2}$, $ \beta \equiv v/c$
  and $X$ is the space-time
  position of the object. Only if the latter is in free space, and undergoes no interaction
  with the force fields of other objects, is the mass, $m$, a constant.
   A simple example of the variability of the masses of  interacting objects
  is provided by the case of two equally massive, equally charged, objects moving towards each
  other along a straight line in their common CM frame. At a certain distance, $x_O$, of
  each object from their common center of energy, they come to rest, due to their mutual repulsion.
  The energy of the system in the CM system (i.e. its mass times $c^2$) is then:
  \begin{equation}
  E = 2 m c^2 +\frac{q^2}{2x_O} \equiv 2 m (x_O) c^2
  \end{equation}
 where $q$ is the charge of each object and $m$ its mass in the absence of interactions.
 The definition of the spatially dependent mass in the last member of (1.7) follows from
  the symmetry of the problem --a system at rest composed of two equal sub-systems at rest.
  Thus:
  \begin{equation}
   m(x_O) = m + \frac{q^2}{4 c^2x_O}
   \end{equation}
 For larger values, $x > x_O$, of the distance, $x$, of each object from the centre of energy, $m(x)$ will
 vary in such a manner as to respect the conservation of relativistic energy:
 \begin{equation}
 E^2 = 4(( m + \frac{q^2}{4 c^2x})^2 c^4+ p^2c^2) \equiv 4(m(x)^2 c^4+ p^2c^2)
  \end{equation}
 Since $E$ is constant, due to the conservative nature of the interaction, it is clear that any
  change in the momentum of the objects must be accompanied by corresponding changes
  in their masses. Another example of position-dependent masses is given below in Section 6 where
  circular Keplerian orbits of a system of two electrically charged objects are discussed.
  \par The structure of this paper is as follows: In the following section the RCED formulae for the 
  electric and magnetic force fields of a uniformly moving electric charge are derived from Coulomb's
   Law and the 4-vector character of
  the electromagnetic potential. In Section 3 small angle Rutherford scattering is calculated
  in different inertial frames, using either RCED force fields or the conventional retarded fields.
    It is shown that the latter, when used in the initial rest frame of an electron scattered
    by a heavy charged particle, give a result incompatible with both the known Rutherford scattering
    formula and the Bethe-Bloch energy-loss relation.
  In this section it is also shown that the forces predicted by the retarded fields do not, in general,
  respect the law of conservation of relativistic transverse momentum. In Section 4 the breakdown
  of Gauss' law for the total flux of the RCED electric force field of a charge in uniform
  motion is demonstrated. In Section 5
  the recent `torque paradox' of Jackson is resolved by employing the RCED force fields, that, unlike
  the conventional retarded ones, are consistent with SR. Section 6 presents an analysis of circular
  Keplerian orbits in RCED. The relativistic generalisation of Kepler's third law for such orbits is
  derived.
  In Section 7 the question raised by 19th century astronomers, and recalled by Eddington~\cite{Eddington},
  concerning
  the possibility of stable orbital motion under the influence of retarded fields, is addressed,
  for the case of circular orbits in electrically bound systems, and a negative answer is found.
  The final section is a summary of the results obtained.  
  
\SECTION{\bf{Force fields of a moving charge in relativistic classical electrodynamics}}
  The expressions for the electric and magnetic fields in RCED found in ~\cite{JHF1} are
  here rederived for the simpler case of a uniformly moving charge by exploiting the 4-vector
 character of the electromagnetic potential, $A$, and assuming Coulomb's law of force between
 static charges with an instantaneous interaction. The usual covariant definitions\footnote{Gaussian
  electromagnetic units are employed.} 
 of the electric and magnetic fields~\footnote{These are derived from first principles
 in ~\cite{JHF1}}\cite{EinsteinGR} are used:
 \begin{eqnarray}
  \vec{E} & = & - \vec{\nabla}A_0-\frac{1}{c}\frac{\partial \vec{A}}{\partial t} \\
  \vec{B} & = & \vec{\nabla} \times  \vec{A}
\end{eqnarray}
 Denoting by S' (coordinates $t'$, $\vec{x}'$) the rest frame of an electric charge $q$ and S 
 (coordinates $t$, $\vec{x}$) a frame, moving with velocity -$v$ along the $x_1'$ axis, in which the
 fields are to be evaluated, then $A' = (q/r; 0,0,0)$ in S'. If $X$, $X'$ are space-time points 
 in S and S' respectively at which the fields are to be evaluated while  $X_q$ and $X_q'$ are
 the corresponding space-time positions of the charge, the quantity $r$ is the
 invariant space-like interval given in S' by\footnote{A time-like metric is used for 4-vector products}:
\begin{equation}
 r^2 = -(X'-X'_q)^2
\end{equation}
 In view of the instantaneous nature of the Coulomb interaction: $t' = t'_q$,
 so that:
\begin{equation}
 r^2 = (\vec{x'}-\vec{x'}_q)^2 
\end{equation}
 The Lorentz transformation equations between the frames S' and S: 
 \begin{eqnarray}
   A_1 & = & \gamma(A_1'+\beta A_0')  \\
  A_0 & = & \gamma(A_0'+\beta A_1')
\end{eqnarray}
give, remembering the Lorentz-scalar character of $r$:
\begin{equation}
    A = (\frac{\gamma q}{r};\frac{\gamma \beta q}{r},0,0)
\end{equation}
 Substituting (2.7) into (2.1) and (2.2) , and performing the partial differentiations,
  noting that, for the partial time derivative in (2.1), $\vec{x}$ is held constant while
   $\vec{x}_q$ is allowed to vary, the following  expressions for the
   RCED electric and magnetic force fields
   are obtained:    
 \begin{eqnarray}
 \vec{E}(RCED)& = & \frac{j_0 \vec{r}}{c r^3}-\frac{\vec{j} (\vec{r} \cdot \vec{v})}{c^2 r^3} \\
  \vec{B}(RCED)& = & \frac{q \gamma (\vec{v} \times \vec{r})}{c r^3} =
  \frac{\vec{j} \times \vec{r}}{c r^3} 
 \end{eqnarray}
 where $j \equiv q u = q(\gamma c; \gamma v,0,0)$ and $u$ is the 4-vector velocity of the charge q.
  Choosing $\vec{r}$ in the 12 plane and with unit vectors $\hat{\imath}$, $\hat{\jmath}$
   parallel to O$x_1$ and  O$x_2$ respectively, (2.8) can be written in the convenient form:
\begin{equation}
 \vec{E}(RCED, S_e) =   \frac{q}{r^2}(\frac{\hat{\imath} \cos \psi}{\gamma} +
      \gamma \hat{\jmath} \sin \psi)
\end{equation}
 where $\cos \psi = (\vec{r} \cdot \vec{v})/|\vec{r}|| \vec{v}|$. This is the RCED formula 
 used in the following section. 
       
\SECTION{\bf{Small angle Rutherford scattering with instantaneous or retarded fields}}
  Rutherford scattering of an electron of charge, $-e$, and mass, $m$, by a nucleus of
  \newline charge, $Ze$, and mass, $M$, is considered in two frames of reference. The first
  is the rest frame, S$_N$
  (coordinates $(t';x',y',z')$) of the nucleus (essentially the overall centre-of-mass frame of the two body scattering problem)
 the second is the rest frame, S$_e$ (coordinates $(t;x,y,z)$), of the initial electron.
 As in a similar discussion in
 \cite{JackRS} a large impact parameter, $b$, is considered so that the
 geometrical deflection of the electron trajectory in S$_N$ and the change in position
 of the electron
 in S$_e$, during the scattering interaction, are neglected. Also neglected are changes in the
 velocities of the electron or nucleus during the scattering interaction.
  Geometrical parameters in the frames S$_N$ and S$_e$ are defined in Figs. 1a  and 1b respectively,
  while Fig.1c shows the definition of the scattering angle, $\theta$, of the electron momentum
 vector in the centre-of-mass frame S$_{CM} \simeq$ S$_N$.
  The scattering angle is given, in the small angle
  approximation that is made here, as :
  \begin{equation}
    \theta \simeq \frac{\Delta p}{p}
  \end{equation}
  $p$ is the electron momentum in  S$_{CM}$ and $\Delta p$ is calculated by integrating the 
  transverse impulse produced by the electric field of the nucleus during the passage of
  the electron in  S$_N$ or of the nucleus in S$_e$. Small effects due to magnetic fields in
   S$_e$ (which, since they are transverse to the electron velocity, do not contribute
   to $|\Delta p|$) are neglected. The electric field is calculated either from the RCED formula (2.10)
  or by use of the standard formula for the electric field of a moving
  charge given in textbooks on classical electrodynamics. This is the same as the `present time'
  version of the retarded Li\'{e}nard-Wiechert (LW)\cite{LW} field, which was first derived by Heaviside
  ~\cite{Heaviside}.
  \par In the frame S$_N$, the electric field at the position of the electron is given, in both
   cases, by a static Coulomb field, since the source charge is at rest:
  \begin{equation}
  \vec{E}(S_N, RCED) = \vec{E}( S_N, LW) =  \hat{\imath'}E_L(S_N) +\hat{\jmath'}E_T(S_N) \equiv
 \frac{Ze}{r^2}(\hat{\imath'} \cos \psi +\hat{\jmath'} \sin \psi)
   \end{equation}
    where $ \psi$ is the angle between the vectors $\vec{r}$ and $\vec{v}$ (see Fig.1a)
    and $\hat{\imath'}$, $\hat{\jmath'}$ are unit vectors in the $x'$- and $y'$-directions. In the frame
    S$_e$, where the nucleus is in motion, different formulae are found for the electric
    field in the two cases:
  \begin{eqnarray}
    \vec{E}(S_e, RCED) & = &   \frac{Ze}{r^2}(\frac{\hat{\imath} \cos \psi}{\gamma} +
      \gamma \hat{\jmath} \sin \psi)   \\
   \vec{E}(S_e, LW) & = & \frac{Ze}{r^2 \gamma^2}\frac{(\hat{\imath} \cos \psi +\hat{\jmath} \sin \psi)}
     {(1-\beta^2 \sin^2 \psi)^{\frac{3}{2}}}
  \end{eqnarray}
 where  $\hat{\imath}$ and $\hat{\jmath}$ are unit vectors in the $x$- and $y$-directions.   
   Although the transverse fields given by (3.3) and (3.4) agree when $\psi = \pi/2$, the angular
   dependence for fixed $r$ is very different, as may be seen in Figs.2 and 3. which show respectively
   $E_Lr^2/Ze$ and $E_T r^2/Ze$, where $E_L$ and $E_T$ are the longitudinal and transverse
 components of $\vec{E}$, repectively, as a function of $\psi$, for various values of $\beta$. It can be seen in Fig.3 that the denominator
   of the right side of (3.4) strongly damps the strength of the transverse electric field, except for
   values of $\psi$ close to $\pi/2$, for values of $\beta$ close to unity. On performing the integral
   over  $\psi$ to calculate $\Delta p$ a much smaller value of $\theta$ is then to be expected for a given 
    value of $b$  for $\vec{E}( S_e, LW)$ as compared to $\vec{E}( S_e, RCED)$. This is indeed found
    to be the case.
\begin{figure}[htbp]
\begin{center}
\hspace*{-0.5cm}\mbox{
\epsfysize15.0cm\epsffile{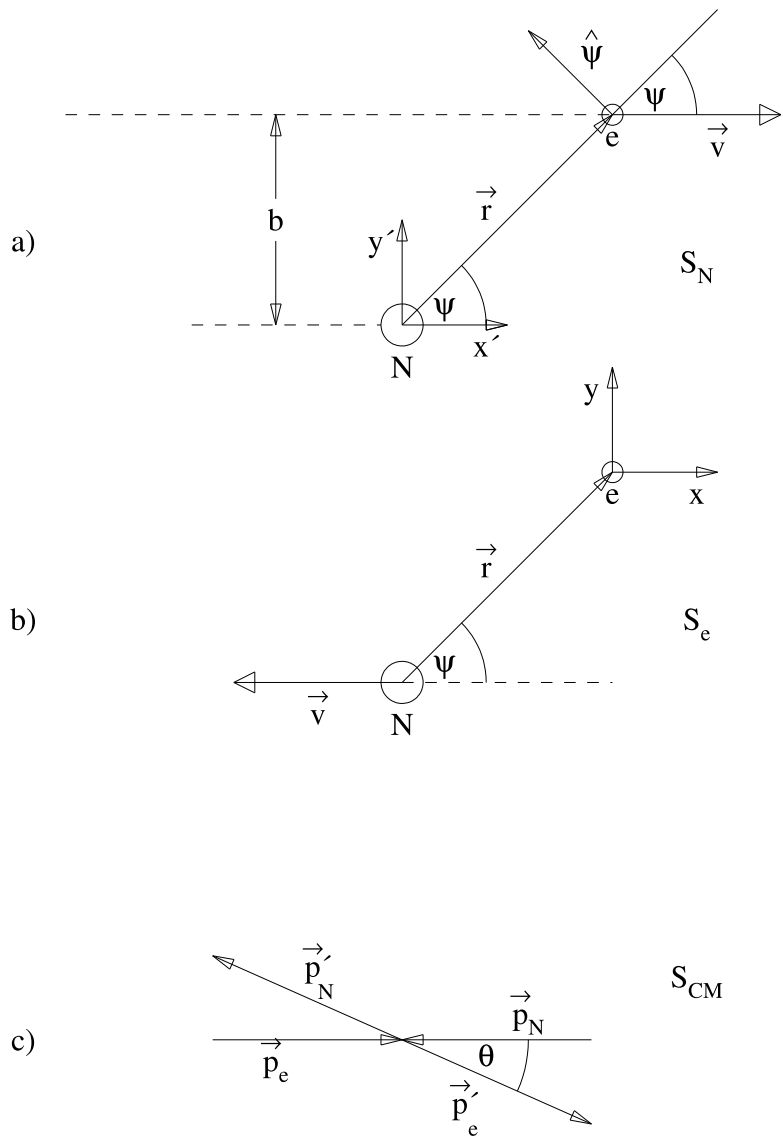}}   
\caption{{\sl Geometrical parameters for the calculation of Rutherford
 scattering in different frames. In a) the nucleus, N, is at rest in S$_N$, and the scattered electron, e,
 is in motion. In b) the electron is initially at rest in S$_e$ while the nucleus is in motion.
 c) shows the definition of the electron scattering angle, $\theta$, in the overall center-of-mass frame,
  S$_{CM}$. To a good approximation, S$_N$ and S$_{CM}$ are the same.}}
\label{fig-fig1}
\end{center}
 \end{figure}

\begin{figure}[htbp]
\begin{center}
\hspace*{-0.5cm}\mbox{
\epsfysize15.0cm\epsffile{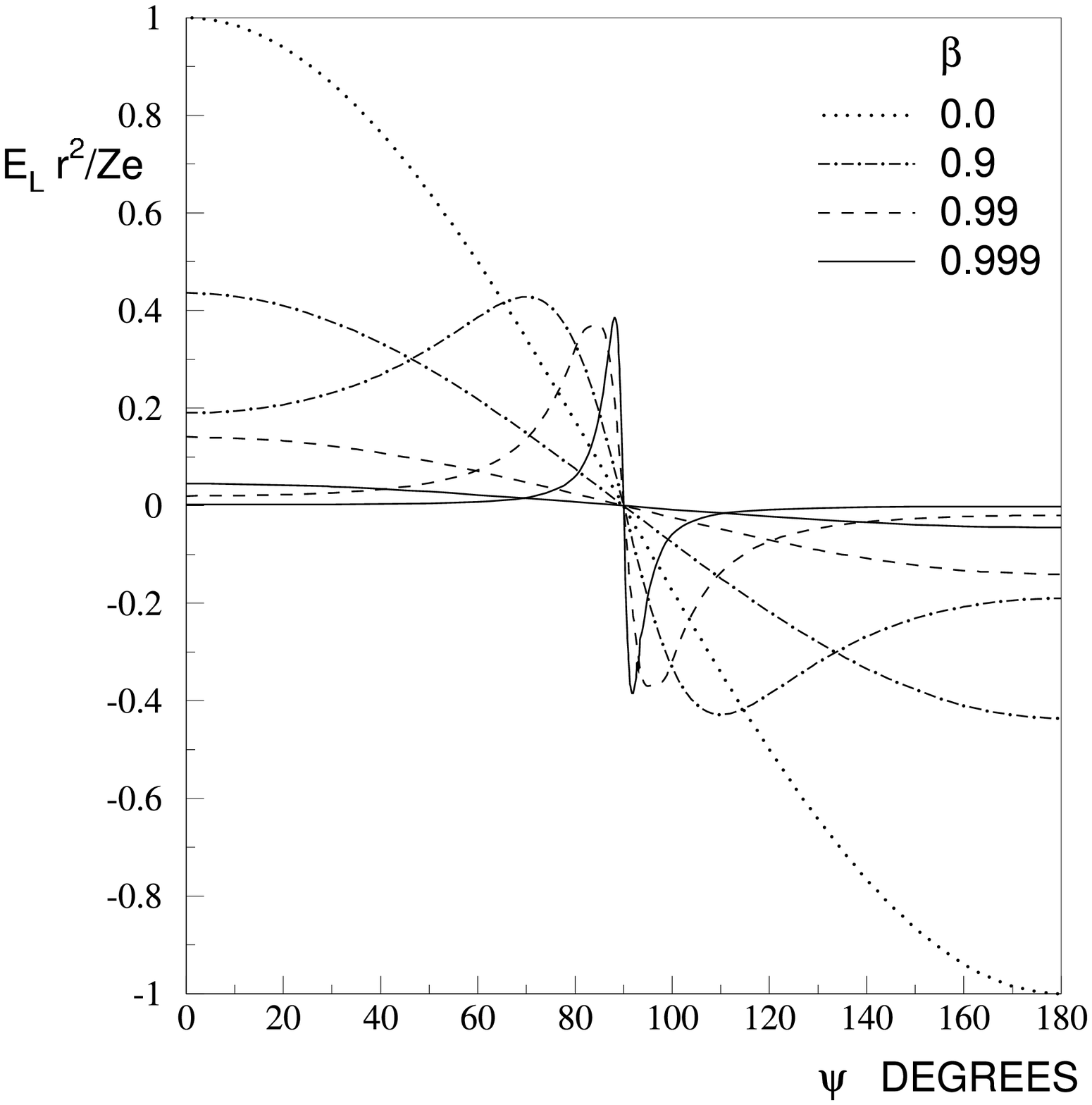}}   
\caption{{\sl The longitudinal electric field of a uniformly moving electric charge, scaled
  by the factor $r^2/Ze$, as a function of angular position $\psi$, for different relativistic
 velocities $\beta = v/c$. The slowly varying cosine curves correspond to the RCED formula (3.3), the more
 rapidly varying ones to the retarded CEM field of (3.4).}}
\label{fig-fig2}
\end{center}
 \end{figure}

\begin{figure}[htbp]
\begin{center}
\hspace*{-0.5cm}\mbox{
\epsfysize15.0cm\epsffile{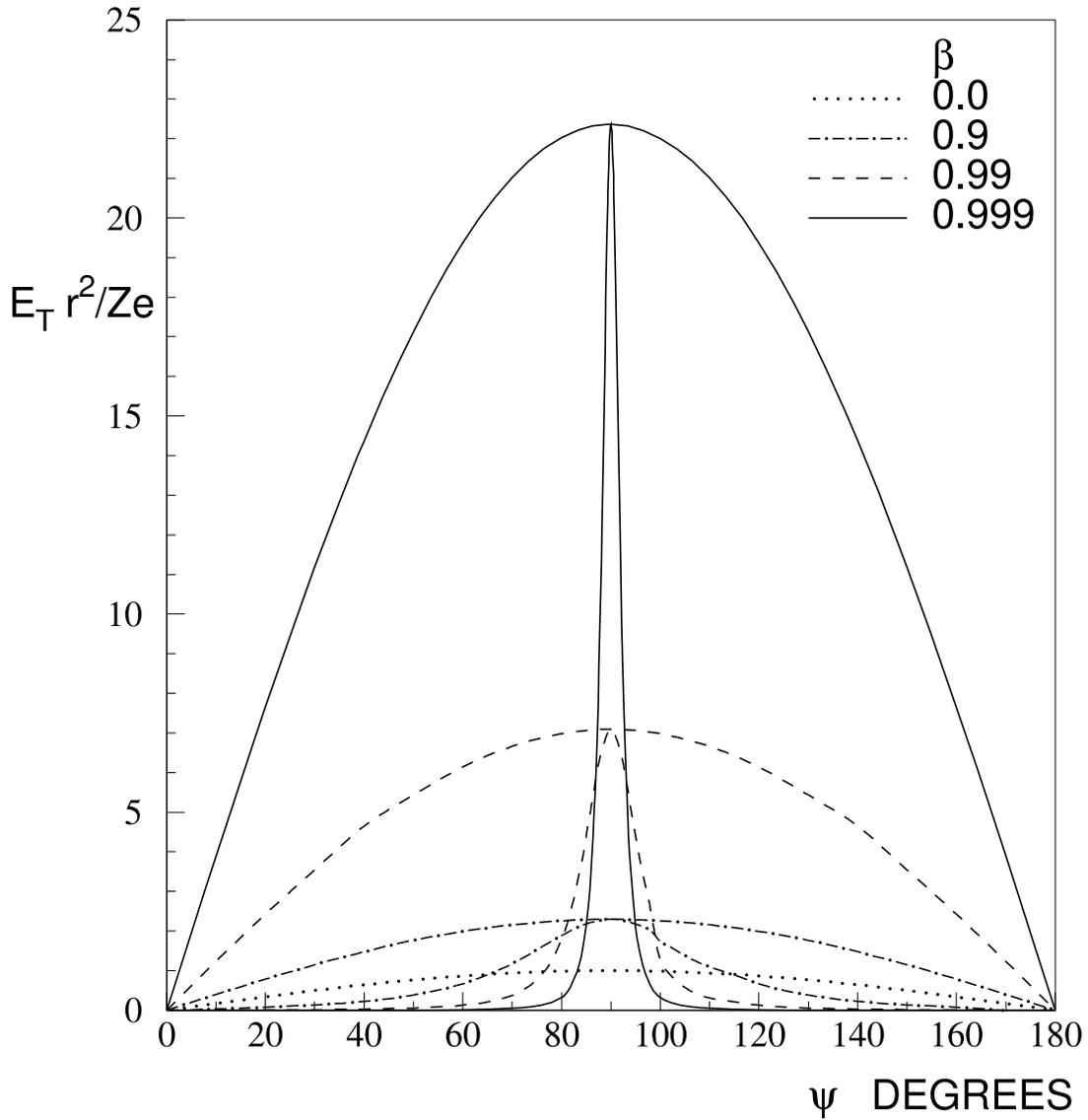}}   
\caption{{\sl The transverse electric field of a uniformly moving electric charge, scaled
  by the factor $r^2/Ze$, as a function of angular position $\psi$, for different relativistic velocities 
  $\beta = v/c$. The slowly varying sine curves correspond to the RCED formula (3.3), the more rapidly 
  varying ones to the retarded CEM field of (3.4).}}
\label{fig-fig3}
\end{center}
 \end{figure}

      \par In the frame S$_N$ the transverse impulse\footnote{$\Delta p$ is treated throughout
    as a positive quantity.}   is given, according to (3.2) as:
      \begin{equation}
      \Delta p(S_N) = \int_{-\infty}^{\infty} eE_T(S_N) dt' =
       Z e^2\int_{-\infty}^{\infty}\frac{\sin \psi}{r^2} dt'
      \end{equation}
       Consideration of the component of the electron velocity $\vec{v}$ perpendicular  to
       $\vec{r}$ in the scattering plane (i.e. in the direction of the unit vector $\hat{\psi}$,
       see Fig.1a) gives 
      the relation: 
      \begin{equation}
         v_{\psi} = r \frac{d \psi}{d t'} = -v \sin \psi
      \end{equation}
       Multiplying both sides of (3.6) by r and noting that, from the geometry of Fig.1a, $ r \sin \psi = b$ gives,
      on rearrangement, 
     \begin{equation}
       d t' = -\frac{r^2 d \psi}{bv}
      \end{equation}
      Using (3.7), (3.5) is converted into an integral over $\psi$:
      \begin{equation}
      \Delta p(S_N) = \frac{Z e^2}{bv}\int_{0}^{\pi}\sin \psi d\psi =  \frac{2 Z e^2}{bv}
      \end{equation} 
     (3.1) and (3.8) give, for the relation between the scattering angle and the impact parameter:
      \begin{equation}
      \theta(S_N)= \frac{2 Z e^2}{p b v}
      \end{equation} 
       in agreement, in the small angle limit, with the exact result (i.e. one valid also for
       $ v \simeq c$)~\cite{JackRS}:
       \begin{equation}
  \tan\frac{\theta}{2} = \frac{Z e^2}{m v^2 b}
      \end{equation} 
    Calculating now the transverse impulse in the frame S$_e$ using the RCED formula (3.3) for the electric
   field gives, instead of (3.5):
       \begin{equation}
      \Delta p(S_e,RCED) =  \int_{-\infty}^{\infty} eE_T(,S_e, RCED) dt = Z e^2\int_{-\infty}^{\infty} \gamma \frac{\sin \psi}{r^2} dt
      \end{equation}    
  Relativistic time dilatation
  gives the relation\footnote{ Note that $t$ is the proper time $\tau$ in the 
  initial electron rest frame, so that $dt' = \gamma d \tau$ is the usual time dilatation
  formula relating time increments in `stationary' and `moving' inertial frames.}: $\gamma  dt = dt'$.
  Hence the right sides of 
 (3.5) and (3.11) are identical,
  leading to the result:
      \begin{equation}
      \theta(S_e, RCED)=  \theta(S_N) = \frac{2 Z e^2}{p b v}
      \end{equation} 
 Using now (3.4) and (3.7) as well as the relation  $\gamma  dt = dt'$,  the transverse impulse imparted
  by the LW electric field in the frame S$_e$ is :
      \begin{equation}
      \Delta p(S_e, LW) =  \int_{-\infty}^{\infty} eE_T(S_e, LW) dt =  \frac{Z e^2}{\gamma^3 bv}\int_{0}^{\pi}\frac{\sin \psi d \psi}
   {(1-\beta^2 \sin^2 \psi)^{\frac{3}{2}}} =  \frac{2 Z e^2}{\gamma b v}
      \end{equation} 
   (3.13) gives, for the electron scattering angle:
       \begin{equation}
      \theta(S_e,LW)= \frac{2 Z e^2}{\gamma p b v}
      \end{equation}
   Thus for a given value of $b$ the value of the scattering angle is reduced by a factor
   $1/\gamma$ as compared to the calculation in S$_N$.
    \par In Jackson's book~\cite{JackRS} a result consistent with (3.12) was found
   for the calculation in the frame S$_e$ using $\vec{E}(S_e, LW)$\footnote{The integration
   variable was a time rather than the angle $\psi$.}. This is because the calculation
   effectively used, in (3.13), the time $t'$ defined in the frame S$_N$ instead of the correct time
   $t$ in the frame
   S$_e$. Making the replacement $dt \rightarrow dt'$ in (3.13) and using (3.7) gives a result, the
   one quoted by Jackson, consistent with (3.12).  A footnote in~\cite{JackRS} indicates that the author
  has realised that $\Delta p$
  was calculated in the frame S$_e$, whereas the relation for the scattering angle, (3.1), 
  is valid, instead, in S$_{CM}$, where the longitudinal electron momentum is $\gamma m v$. There is then the
   statement: `The reader may verify that (13.3)' (i.e. (3.9) above)`and (13.4)'(i.e. (3.10) above)
  'are also correct in the frame in which the electron is at rest by transforming the angles from
  the CM system to the laboratory'. Actually, this is impossible to do, because the electron
  scattering angle is undefined in the electron rest frame. The calculations of $\Delta p$
  in the frames S$_N$ and S$_e$ above using the retarded field (3.4) show, instead, results
   that differ by a factor $\gamma$. The calculation of the scattering angle performed in
    S$_e$ then gives an angle a factor $\gamma$ smaller than the known Rutherford scattering 
    result (2.10). Use of  $\Delta p$ calculated
    with (3.4) in  S$_e$ then leads to a $(1/\gamma)^2$ dependence of the energy-loss formula
    at relativistic velocities, in contradiction both with the Bethe-Bloch formula and
   experiment~\cite{PDGIL}.
   \par That the transverse field in (3.3) is larger, by a factor $\gamma$, than the static value in (3.2),
    for any value of $\psi$, is a necessary consequence of the invariance of relativistic transverse
    momentum will now be demonstrated. Since the Lorentz transformation leaves invariant transverse
    spatial intervals:
    $\Delta y = \Delta y'$, relativistic transverse momentum must also be conserved:
   \begin{equation}
    p_T \equiv m \frac{dy}{d \tau} = m \frac{dy'}{d \tau} =  m \gamma  v = m \gamma'  v'
   \end{equation}
  Considering now the transverse momentum, $p_T$, of the electron in Rutherford scattering 
  and associating the frame S with S$_{N}$ and S' with S$_e$\footnote{This, the opposite of the previous 
  assignment, is done for clarity of presentation. Here it is the rest and moving frames of the
  scattered electron rather than those of the nucleus $N$ that are
  under discussion.}, then, since
  $v' \ll v$, $\gamma' \simeq 1$. 
  That is, the motion of the electron in  S$_e$ is non-relativistic. If $p_T$ is generated by transverse
  forces $F_T$, $F_T'$ in  S$_{N}$, S$_e$ respectively, conservation of transverse momentum gives:
   \begin{equation}
  \Delta p_T = F_T \Delta t =  F_T' \Delta t'
  \end{equation}
  so that
   \begin{equation}
  \frac{ F_T'}{ F_T} =   \frac{ \Delta t}{\Delta t'} =   \frac{\gamma}{\gamma'} \simeq \gamma 
   \end{equation}
    since $\Delta \tau = \Delta t/\gamma =  \Delta t'/ \gamma'$.
   Thus conservation of transverse momentum requires the transverse electric field in  S$_e$
   to be a factor $\gamma$ larger than that in  S$_{N}$ for all values of $\psi$. This condition
   is respected by $\vec{E}(RCED)$ in (3.3), but not by  $\vec{E}(LW)$ in (3.4), for which the condition
   is satisfied only for $\psi = \pi/2$. The forces generated by the retarded field (3.4) therefore do not, in
  general, respect the invariance of relativistic transverse momentum, in different
  inertial frames, that is required by SR.

\SECTION{\bf{Breakdown of Gauss' theorem for the electric field flux of a moving charge
  in relativistic classical electrodynamics}}
 The total flux of the electric field generated by a moving charge:
    \begin{equation}
  \phi_E  = \int_S \vec{E} \cdot d \vec{S}
 \end{equation}
   is readily calculated from the formulae (3.3) and (3.4) for different choices of a surface, S,
  that completely surrounds the charge. The results obtained for (i) a spherical surface centered
  on the charge and (ii) a cylindrical surface with axis parallel to the direction of motion of the charge,
  of length $2 \ell$ and radius $a$, with the charge at the axis mid-point, are, for the RCED
  and the retarded LW fields:
   \newline
   \newline
   Sphere:
 \begin{eqnarray}
\phi_E(RCED) &  = &  \frac{4 \pi Z e}{3}\left(2 \gamma+ \frac{1}{\gamma}\right) \\
\phi_E(LW) &  = & 4 \pi Z e  
 \end{eqnarray}
   \newline
   Cylinder:
 \begin{eqnarray}
\phi_E(RCED) &  = &  4 \pi Z e\left((\gamma-\frac{1}{\gamma})cos \theta_0 + \frac{1}{\gamma}\right) \\
\phi_E(LW) &  = & 4 \pi Z e  
 \end{eqnarray}
 where $\tan \theta_0 = a/\ell$.
 \par The flux of the retarded LW field is independent of the velocity of the 
  charge and of the form of the surface S, but, as can be seen in Fig.3, the electric field
  strength peaks strongly near
  $\psi = \pi/2$ for large velocities $\beta \simeq 1$. The number of field lines is conserved
 but their spatial distribution changes with the velocity of the charge. This change in the
  distribution of the field lines is often interpreted as due to relativistic `length contraction'
  ~\cite{JackLC}. However, the flux of the RCED force fields do not, as is clear from (4.2) and (4.4),
  respect Gauss' theorem, or the electrostatic Maxwell equation, when the source charge is 
  in motion. The flux depends both on the velocity of charge and the geometry of the surface S.
  For large values of $\gamma$ the flux is proportional to $\gamma$, but with different 
  multiplicative constants for spherical or cylindrical surfaces. In RCED, both the tranverse
  field and the total flux are proportional to $\gamma$ at large $\gamma$, whereas in the same limit
  the transverse LW field is proportional to  $\gamma$ only for  $\psi = \pi/2$, while the total flux
  is velocity independent. There should be nothing shocking in these properties of the
  fields. As discussed in ~\cite{JHF1}, the force fields are only mathematical abstractions
  introduced in order to write inter-particle forces in a compact manner. The physical reality
  resides in the particles and the forces between them, not in the fields. All equations of
  motion may be written down and solved without the necesssity to
 introduce either the concept of a force
  or that of a field. Just this approach is taken in the discussion of Keplerian orbits
  in Section 6 below. The total flux of the field, though readily calculable, has no independent
  physical meaning and it is irrelevant, for the dynamical description of the motion of the
  charges, whether the flux respects, of not, Gauss' theorem. However the forces derived from
  the fields must respect on the one hand, the constraints of SR and on other be consistent,
  in the appropraiate limit, with the known and tested predictions of
  QED. As the several examples considered in the present paper show, this is the case for the instantaneous
  RCED fields, but not for the retarded LW ones.

 \SECTION{\bf{Resolution of Jackson's torque paradox}}

 Another example of the incompatibility of the LW retarded fields with relativistic
 invariance is provided by the `torque paradox' recently pointed out by Jackson~\cite{JackTP}.
 Replace in Fig.1b the  moving nucleus by an electron at rest. The electric field at $e$
 is then a static Coulomb field parallel to $\vec{r}$. This is the only force-generating field
 in the frame S$_e$ and gives no torque. Now, on boosting from S$_e$ into S$_N$, both electrons
 move with velocity $v$ parallel to the x-axis. According to the LW formula (3.4), the electric 
  field is still parallel to $\vec{r}$ and therefore generates no torque. However, in S$_N$,
  the electric current constituted by the lower electron produces at $e$ a magnetic field
  of magnitude:
   \begin{equation}
  \vec{B}(LW) = \vec{\beta} \times \vec{E}(LW)
   \end{equation}
  The corresponding Lorentz force on $e$ is parallel to Oy and gives a non-vanishing
  torque, $\vec{T}$:
    \begin{equation}
    \vec{T} = -e \vec{r} \times [\vec{\beta} \times (\vec{\beta} \times \vec{E}(LW))]
      = -\hat{k} e \beta^2 \sin \psi  \cos \psi |\vec{E}(LW)|  
     \end{equation} 
   where $\hat{k}$ is a unit vector parallel to the
   $z$-axis\footnote{This formula is identical to Eqn(7)
   of~\cite{JackTP} on making the replacements $\theta  \rightarrow
   \psi$, $\beta_0  \rightarrow \beta$ and $t = 0$ in the latter.}.
   Since this torque does not exist in the inertial frame S$_e$, related to S$_N$ by
   Lorentz transformation, there is a manifest breakdown of the relativity principle.
   This is Jackson's torque paradox. A similar problem was previously pointed out ~\cite{PPTN}
  in connection with the interpretation of Trouton and Noble experiment~\cite{TN}.                  
   \par The RCED fields at $e$ in S$_N$, on replacing N by an electron moving with velocity $v$ along
   Ox' are, using (2.9) and (2.10), and dropping the primes on the unit vectors in S$_N$:
   \begin{eqnarray}
    \vec{E}(RCED) & = &  -\frac{e}{r^2}(\frac{\hat{\imath} \cos \psi}{\gamma} +
      \gamma \hat{\jmath} \sin \psi)   \\
  \vec{B}(RCED)   & = & -\frac{e \hat{k}}{r^2}\gamma \beta \sin \psi
  \end{eqnarray}  
   The corresponding Lorentz force, $\vec{F}$, on $e$ is:
 \begin{eqnarray}
     \vec{F}(RCED) & = & -e [ \vec{E}(RCED) + \vec{\beta} \times \vec{B}(RCED)] \nonumber \\
       & = & \frac{e^2}{r^2}\left[\frac{\hat{\imath} \cos \psi}{\gamma} +
      \gamma \hat{\jmath} \sin \psi + \beta (\hat{\imath} \times \hat{k})(\gamma \beta \sin \psi)\right]
 \nonumber \\ 
        & = & \frac{e^2}{r^2}\left[\frac{\hat{\imath} \cos \psi}{\gamma} + \hat{\jmath} \gamma (1-\beta^2) 
      \sin \psi \right]
 \nonumber \\ 
        & = & \frac{e^2}{\gamma r^2}\left[\hat{\imath} \cos \psi + \hat{\jmath} \sin \psi \right]
   = \frac{e^2}{\gamma r^2} \vec{r} 
  \end{eqnarray}  
 Thus the Lorentz force in S$_N$ is radial and, consistent with relativistic invariance, there is no torque
  in this frame.

 \SECTION{\bf{Circular Keplerian orbits in Relativistic Classical Electrodynamics}}
  In ~\cite{JHF1} equations of motion, of two mutually interacting charged objects
  O$_1$ and O$_2$, were derived. 
  They may be most simply written in the `fieldless' and `forceless' form:
    \begin{eqnarray} 
      \frac{d\vec{p_1}}{dt} &  = & m_1 \frac{d(\gamma_1 \vec{v_1})}{d t} =
     \frac{q_1}{c}\left[\frac{ j_2^0\vec{r} +  \vec{\beta}_1 \times
     (\vec{j_2} \times \vec{r})}{r^3} -\frac{1}{c r}\frac{d \vec{j_2}}{d t}-\vec{j_2}
     \frac{(\vec{r} \cdot \vec{\beta}_2)}{r^3} 
      \right] \\
       \frac{d\vec{p_2}}{dt}  &  = &  m_2 \frac{d(\gamma_2 \vec{v_2})}{d t} =
 -\frac{q_2}{c}\left[\frac{ j_1^0\vec{r} +\vec{\beta}_2 \times 
  (\vec{j_1} \times \vec{r})}{r^3}+\frac{1}{c r}\frac{d \vec{j_1
}}{d t} -\vec{j_1}
     \frac{(\vec{r} \cdot \vec{\beta}_1)}{r^3}
    \right] 
     \end{eqnarray}
    where $m_1$ and $m_2$ are the masses of the objects in the absence of mutual interaction,
    and $q_1$ and $q_2$ their electric charges.
    The vector $\vec{r} \equiv \vec{r_{12}}\equiv \vec{r_1}-\vec{r_2}$ gives the spatial separation 
   of the objects in their centre-of-mass  frame. The 4-vector current $j_i$ of the object
    O$_i$ is defined as $j_i \equiv q_i u_i$, where $u_i$ is the 4-vector velocity 
    $u_i \equiv (c\gamma_i;c\gamma_i \vec{\beta_i})$.  In the following, the motion of the objects
    in their centre-of-mass frame is considered, and the origin of the vectors $\vec{r_1}$ and  $\vec{r_2}$
    is the common centre of energy, O, of the objects.
    \par Because of the relation connecting the velocity, current and energy-momentum 4-vectors:
   \begin{equation}
       u = \frac{j}{q}= \frac{p}{m}
   \end{equation}
    the differential equations (6.1) and (6.2) are coupled via the $d \vec{j}/dt$ terms on their
    right sides. Solving these equations for the time derivatives of the relativistic velocities
    of the two objects gives the equations:
     \begin{eqnarray} 
     \frac{d(\gamma_1 \vec{v}_1)}{d t} &  =  & \frac{q_1 q_2}{m_1\left[1-\frac{(q_1 q_2)^2}{m_1 m_2 c^4
      r^2}\right]} \frac{1}{r^3} \times \nonumber \\
         &   & \left\{ \gamma_2[\vec{r}+ \vec{\beta}_1 \times ( \vec{\beta}_2 \times \vec{r})
          -  \vec{\beta}_2 (\vec{r} \cdot \vec{\beta}_2)] \right. \\
     &   & \left. +\frac{q_1 q_2 \gamma_1}{m_2 c^2 r}[\vec{r}+ \vec{\beta}_2 \times ( \vec{\beta}_1
      \times \vec{r}) -  \vec{\beta}_1 (\vec{r} \cdot \vec{\beta}_1)] \right\} \nonumber \\
    \frac{d(\gamma_2 \vec{v_2})}{d t} &  =  & \frac{- q_1 q_2}{m_1\left[1-\frac{(q_1 q_2)^2}{m_1 m_2 c^4
      r^2}\right]} \frac{1}{r^3} \times \nonumber \\
         &   & \left\{ \gamma_1[\vec{r}+ \vec{\beta}_2 \times ( \vec{\beta}_1 \times \vec{r})
          -  \vec{\beta}_1 (\vec{r} \cdot \vec{\beta}_1)] \right.\nonumber \\
     &   & \left. +\frac{q_1 q_2 \gamma_2}{m_1 c^2 r}[\vec{r}+ \vec{\beta}_1 \times ( \vec{\beta}_2
      \times \vec{r}) -  \vec{\beta}_2 (\vec{r} \cdot \vec{\beta}_2)] \right\}
     \end{eqnarray}
  \par The differential geometry of uniform circular motion is illustrated in Fig.4. The vectors 
    $\vec{v}_1$ or $\vec{v}_2$ are perpendicular to $\vec{r}$ at all times. Also $|\vec{v}_1|$,
      $|\vec{v}_2|$, $r_1$ and $r_2$ are constant at all times. It follows from the geometry of
       Fig.4. that, to first order in $\delta r$,$\delta v$ :
      \begin{equation}
   \frac{\delta \phi}{\delta t}= \frac{1}{v_1}\frac{|\delta \vec{v}_1|}{\delta t} = 
   \frac{1}{v}_2\frac{|\delta \vec{v}_2|}{\delta t}
      = \frac{1}{r}_1\frac{|\delta \vec{r}_1|}{\delta t} = \frac{v_1}{r_1}              
     = \frac{1}{r_2}\frac{|\delta \vec{r}_2|}{\delta t} = \frac{v_2}{r_2}
   \end{equation}
    For the case of uniform circular motion, $q_1$ and $q_2$ must have opposite
    signs and (6.4) and (6.5) simplify to:

     \begin{eqnarray} 
   \gamma_1  \frac{d \vec{v}_1}{d t} &  =  & \frac{-|q_1||q_2|(1+\beta_1 \beta_2)}{m_1\left[1
    -\frac{(q_1 q_2)^2}{m_1 m_2 c^4  r^2}\right]}\left [\gamma_2-\frac{|q_1||q_2| \gamma_1}{m_2 c^2 r}
      \right] \frac{\vec{r}}{r^3} \\  
  \gamma_2  \frac{d \vec{v}_2}{d t} &  =  & \frac{|q_1||q_2|(1+\beta_1 \beta_2)}{m_2\left[1
    -\frac{(q_1 q_2)^2}{m_1 m_2 c^4  r^2}\right]}\left [\gamma_1-\frac{|q_1||q_2| \gamma_2}{m_2 c^2 r}
      \right] \frac{\vec{r}}{r^3}         
      \end{eqnarray}
    These equations may be combined to yield the relation:
   \begin{equation}   
    {\cal E}_1^* \frac{d\vec{v}_1}{d t}+ {\cal E} _2^*\frac{d\vec{v}_2}{d t} = 0
  \end{equation}
    where 
      \begin{eqnarray} 
 {\cal E}_1^* & \equiv & \frac{\gamma_1 m_1}{\gamma_2-\frac{|q_1||q_2| \gamma_1}{m_2 c^2 r}} \\
  {\cal E}_2^* & \equiv & \frac{\gamma_2 m_2}{\gamma_1-\frac{|q_1||q_2| \gamma_2}{m_1 c^2 r}} 
      \end{eqnarray}
   Since, for uniform circular motion, ${\cal E}_1^*$ and ${\cal E}_2^*$ are constant,
   (6.9) may also be written as:
    \begin{equation}   
   \frac{d }{d t} [{\cal E}_1^* \vec{v}_1+{\cal E}_2^* \vec{v}_2] =  
    \frac{d }{d t}[\vec{p_1^*} + \vec{p_2^*}] = 0 
  \end{equation} 
   or, equivalently,
     \begin{equation}   
   \frac{d \vec{p_1^*}}{d t}  = -\frac{d \vec{p_2^*}}{d t}   
     \end{equation}
     where $\vec{p_i^*} \equiv {\cal E}_i^* \vec{v}_i$,  $i=1,2$. This is just the expression
     of Newton's third law for the two mutually interacting objects.       
    \par  Equations (6.10) and (6.11) then suggest that the electromagnetic interaction between
  the objects modifies their masses as a function of their velocites and spatial separation according
   to the relations:
      \begin{eqnarray} 
 m_1^* & \equiv & \frac{m_1}{\gamma_2-\frac{|q_1||q_2| \gamma_1}{m_2 c^2 r}} \\
  m_2^* & \equiv & \frac{m_2}{\gamma_1-\frac{|q_1||q_2| \gamma_2}{m_1 c^2 r}} 
      \end{eqnarray} 
   Thus, when proper account is taken of the modification of the masses of the objects,
   due to their mutual electromagnetic interaction, there is not, in the case 
   considered here, any breakdown of Newton's third law.
\begin{figure}[htbp]
\begin{center}
\hspace*{-0.5cm}\mbox{
\epsfysize12.0cm\epsffile{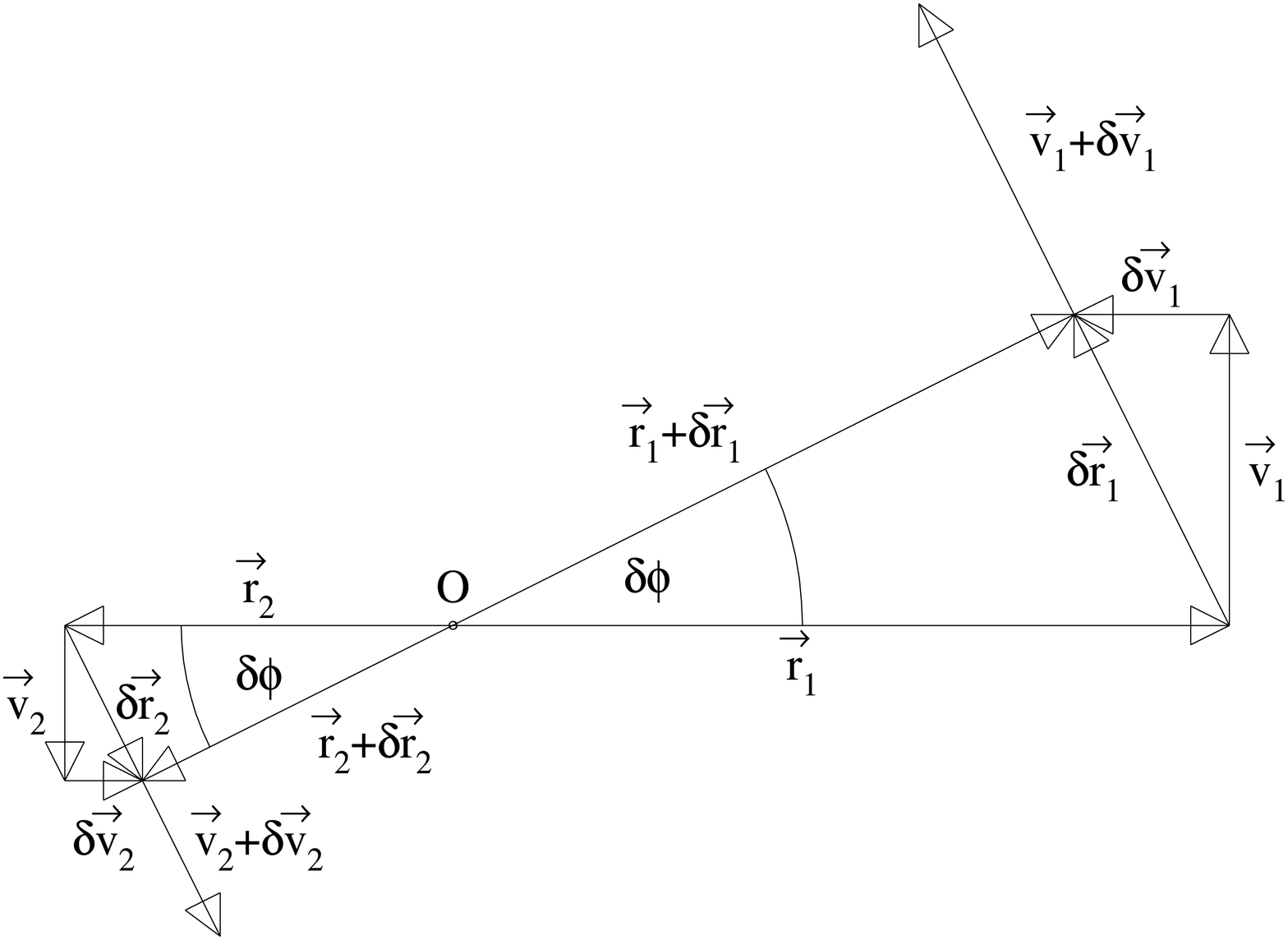}}   
\caption{{\sl  Differential geometry of the position and velocity vectors of two objects, O$_i$, $(i =1,2)$
 executing uniform circular motion about their common center-of-energy O. As $\delta \vec{r}_i$ and 
  $\delta \vec{v}_i$  $\rightarrow 0$ , $\delta \vec{r}_i \bot \vec{r}_i$,  $\delta \vec{v}_i \bot \vec{v}_i$,
  $|\vec{r}_i| = $ const. and  $|\vec{v}_i| = $ const.}}
\label{fig-fig4}
\end{center}
 \end{figure}   

   \par Combining (6.6) and (6.9)
     \begin{equation}
    \frac{|d\vec{v}_1|/dt}{|d\vec{v}_2|/dt} = \frac{{\cal E}_2^*}{{\cal E}_1^*} =\frac{v_1}{v_2}
   = \frac{r_1}{r_2}
     \end{equation}
    or
      \begin{eqnarray} 
{\cal E}_2^* v_2 & = & {\cal E}_1^* v_1 \\
{\cal E}_2^* r_2 & = & {\cal E}_1^* r_1
     \end{eqnarray} 
  Equation(6.17), which may be written as $|\vec{p_2^*}| = |\vec{p_1^*}|$, is consistent 
   with momentum conservation in the overall centre-of-mass frame:
   $\vec{p_2^*} = -\vec{p_1^*}$, while (6.18) states that the origins of $\vec{r}_1$
   and $\vec{r}_2$ are at the centre of energy of the two interacting objects. 
    \par The period of rotation, $\tau$, of either object in its circular Keplerian orbit, 
   is readily derived from (6.7) or (6.8) and the geometrical relation (6.6).
    Considering the object O$_1$:    
     \begin{eqnarray} 
   \gamma_1  \frac{|d \vec{v_1}|}{d t} &  =  & \frac{|q_1||q_2|(1+\beta_1 \beta_2)}{m_1 \gamma_1 \left[1
    -\frac{(q_1 q_2)^2}{m_1 m_2 c^4  r^2}\right]}\left [\gamma_2-\frac{|q_1||q_2| \gamma_1}{m_2 c^2 r}
      \right] \frac{1}{r^2} =  v_1 \frac{d \phi}{d t} = \frac{2 \pi r_1}{\tau} \frac{2 \pi}{\tau}
   \end{eqnarray}
    Using the relations:
   \begin{equation}
    r = r_1+r_2 = r_1(1+\frac{r_2}{r_1})= r_1(1+\frac{{\cal E}_1^*}{{\cal E}_2^*})
     = \frac{r_1{\cal E}_1^*}{{\cal E}^*}
      \end{equation}
     where ${\cal E}^*$ is the `reduced energy':
    \begin{equation}
    {\cal E}^* \equiv \frac{{\cal E}_1^*{\cal E}_2^*}{{\cal E}_1^*+{\cal E}_2^*}  
       \end{equation}
 to eliminate $r_1$ in favour of $r$ on the right side of (6.19) yields a
  formula relating the rotation period of the objects to their spatial separation:
   \begin{eqnarray} 
 \tau^2 & = & \frac{(2 \pi)^2 m_1 \gamma_1 {\cal E}^* \left[1 -\frac{(q_1 q_2)^2}{m_1 m_2 c^4 r^2}\right]
  r^3}
   {{\cal E}_1^*|q_1||q_2|(1+\beta_1 \beta_2)\left[\gamma_2-\frac{|q_1||q_2| \gamma_1}{m_2 c^2 r}\right]}
    \nonumber \\
        & = & \frac{(2 \pi)^2 {\cal E}^* \left[1 -\frac{(q_1 q_2)^2}{m_1 m_2 c^4  r^2}\right] r^3}
   {|q_1||q_2|(1+\beta_1 \beta_2)}
   \end{eqnarray}
    where, in the last member of (6.22), (6.10) has been used. This is the relativistic generalisation 
    of Kepler's third law. The usual classical result is recovered in the limit
    $c \rightarrow \infty$. Note that no kinematial approximation whatever has been made in the 
   derivation. Equation (6.22) gives the exact result for the period of a circular orbit to all orders in
    $|q_1||q_2|/(m_1c^2 r)$, $|q_1||q_2|/(m_2c^2 r)$, $\beta_1$ and $\beta_2$.

\SECTION{\bf{The impossibility of uniform circular motion under retarded Li\'{e}ard-Wiechert fields}}
  Several years before the advent of special relativity the effect on electric and magnetic fields, 
  propagated at the speed of light, was considered by Li\'{e}nard and Wiechert~\cite{LW}. A good explanation
  of the physics arguments presented by these authors may be found in Reference~\cite{PP}. In this case the values of the fields
  at a given space-time point are given by the sum of the contributions from all sources that lie on the
  backward three-dimensional light-cone of the space-time point considered. Neglecting relativistic 
  effects (as of course  Li\'{e}nard and Wiechert were obliged to do) the instantaneous classical scalar and vector
  potentials $\phi(X) = q/r$ and $\vec{A}(X) = q \vec{v}/r$ are found to be modified  according to the relations\cite{PP}:

      \begin{equation}
      \phi_{ret}(X) = \left\{\frac{q}{(r-\frac{\vec{v} \cdot \vec{r}}{c})}\right\}
       _{t_{ret}}
       \end{equation}
      \begin{equation}
      \vec{A}_{ret}(X) = \left\{\frac{q\vec{v}}{c(r-\frac{\vec{v} \cdot \vec{r}}{c})}\right\}
       _{t_{ret}}
       \end{equation}
  where the quantities in large curly brackets are evaluated at the retarded time: $t_{ret} = t-r/c$ at which the
  source charge is sampled by a signal moving along the backward
  light-cone of the space-time point
  $X = (ct;\vec{x})$. 
   The electric and magnetic fields derived from these potentials are given by the expressions\cite{LLCFLW}:
  \begin{eqnarray}
     \vec{E}_{ret}^{vel}(X) & = & \left\{\frac{q(1-\beta^2)(\vec{r}-\vec{\beta}r)}{(r-\vec{\beta}
  \cdot \vec{r})^3}\right\}_{t_{ret}} \\
    \vec{E}_{ret}^{acc}(X) & =  & \left\{\frac{q \vec{r} \times [(\vec{r}-\vec{\beta}r)
  \times d\vec{\beta}/dt]}{c(r-\vec{\beta} \cdot \vec{r})^3}\right\}_{t_{ret}} \\
    \vec{B}_{ret}(X) & =  & -(\vec{E}_{ret}^{vel}(x)+ \vec{E}_{ret}^{acc}(x)) \times \left\{
  \frac{\vec{r}}{r}\right\}_{t_{ret}}
    \end{eqnarray}
   $\vec{E}_{ret}^{vel}(X)$ and $ \vec{E}_{ret}^{acc}(X)$ give the contributions to the electric field
    due to the velocity and acceleration respectively  of the source charge.

\begin{figure}[htbp]
\begin{center}
\hspace*{-0.5cm}\mbox{
\epsfysize15.0cm\epsffile{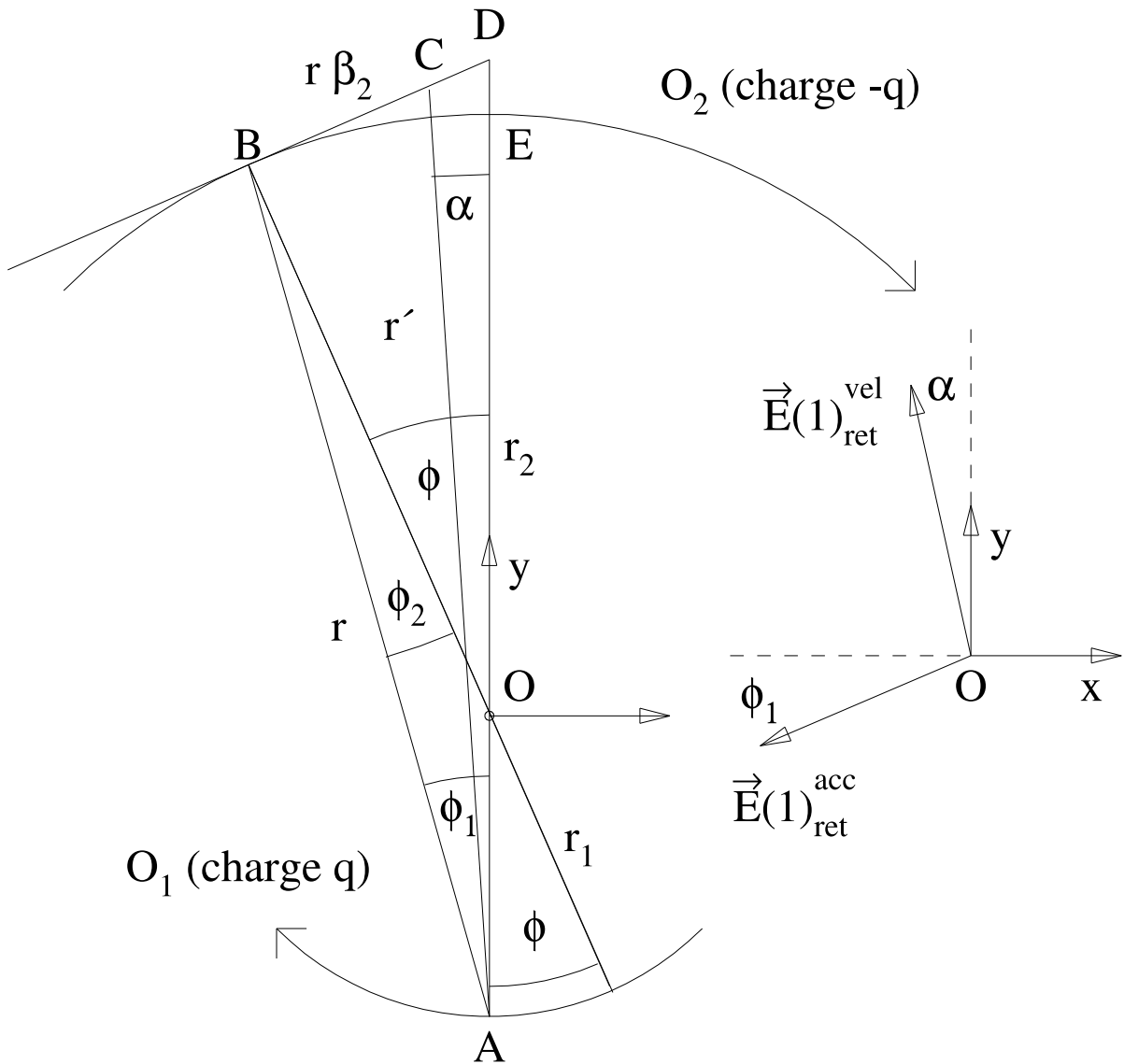}}   
\caption{{\sl Geometry of uniform circular motion of two oppositely charged objects, O$_1$ and O$_2$,
  about their common centre of energy O. The forces on O$_1$ at A generated by the retarded fields
 (7.3)-(7.5) of O$_2$ are calculated. The force fields are produced by O$_2$ at the position B.
  Various geometrical parameters are defined. The directions of the electric fields at A associated with
  the velocity and acceleration of O$_2$ are shown in the inset figure.}}
\label{fig-fig5}
\end{center}
 \end{figure}
   
 \par The possibility of uniform circular motion of two objects with equal and opposite electric charges and
   different masses, interacting mutually via the fields (7.3)-(7.5) above is now investigated. As in the
   previous 
   section, the geometrical constraints of uniform circular motion are imposed and a solution of the equations
   of motion of the objects consistent with these constraints is sought. Various geometrical parameters used 
   to describe the system are defined in Fig.5. The objects O$_1$ and O$_2$ of masses, in the 
   absence of interaction, $m_1$ and $m_2$ and electric
   charges $q$ and $-q$ are assumed to perform uniform circular motion about their centre of energy O.
   The radii of the circular orbits of  O$_1$ and O$_2$ are $r_1$ and $r_2$. Cartesian axes Ox, Oy and Oz
  specified by unit vectors $\hat{\imath}$, $\hat{\jmath}$ and $\hat{k}$ are chosen with y-axis parallel
  to $\vec{r}_2$
  at some instant and x-axis in the plane of the orbits. The resultant force on O$_1$ due to the retarded
   fields of  O$_2$ is now calculated and examined for consistency with uniform circular motion.
   According to Eqn(7.3) $\vec{E}_{ret}^{vel}(X_1) \equiv \vec{E}_{ret}^{vel}(1)$
   points to the `present position' (assuming uniform motion)
   of the source charge. Calculating the retarded fields of  O$_2$ at O$_1$ when the latter is at position
  A in Fig.5 at $t=0$, implies that the fields are given by the formulae (7.3)-(7.5) when  O$_2$ was at
  position B at time $t= -r/c$. $\vec{E}_{ret}^{vel}(1)$ points towards the point C on the tangent to the
  orbit of O$_2$ at B. The distance BC is $r\beta_2$\footnote{ The object O$_2$ does not actually follow this
   trajectory under its assumed circular orbit. However, causality implies that only the velocity 
  of the source at the instant it is sampled on the backward light-cone of the point at which the 
   field is evaluated is relevant. Any change in the motion (velocity or acceleration) of the source
  at later times therefore cannot affect the retarded field. The retarded field $\vec{E}_{ret}^{vel}(1)$ at A
  is therefore the same whether O$_2$ follows a circular trajectory BE or the straight one BC.}. The calculation
  of the retarded fields from the Eqns(7.3)-(7.5) and the geometry of Fig.5 is presented in the Appendix. The
  following results are obtained:
  \begin{eqnarray}
   \vec{E}_{ret}^{vel}(1) & = &
   -\frac{\hat{\imath}q \beta_2^3}{3r_2^2}+\frac{\hat{\jmath}q[1-\beta_2(2\beta_1+\beta_2/2)]}{(r_1+r_2)^2} +O(\beta_2^4) \\
  \vec{E}_{ret}^{acc}(1) & = & -\frac{\hat{\imath} q \beta_2^2(\beta_1+ \beta_2)}{r_2(r_1+r_2)}
    +O(\beta_2^4) \\
   \vec{B}_{ret}(1) & = & -\frac{\hat{k} q \beta_1}{(r_1+r_2)^2} +O(\beta_1^3)  
\end{eqnarray}
    According to the Lorentz force equation, the total force acting on O$_1$ is:
  \begin{equation}
  \vec{F}_{ret}(1) = -\frac{\hat{\imath} q^2 \beta_2^2}{r_2}\left[ \frac{\beta_1 +\beta_2}{r_1+r_2}+\frac{\beta_2}{3 r_2}\right]
  +\frac{\hat{\jmath} q^2}{r_1+r_2}\left[\frac{1-\beta_2(2\beta_1+\beta_2/2)}{r_1+r_2}
    \right]+O(\beta_2^4) 
   \end{equation}
   By symmetry, the force on O$_2$ due to the retarded fields of O$_1$ is:
  \begin{equation}
  \vec{F}_{ret}(2) = \frac{\hat{\imath}q^2 \beta_1^2}{r_1}\left[\frac{\beta_1 +\beta_2}{r_1+r_2}+\frac{\beta_1}{3 r_1}\right]
  -\frac{\hat{\jmath}q^2}{r_1+r_2}\left[\frac{1-\beta_1(2\beta_2+\beta_1/2)}{r_1+r_2}\right]+O(\beta_1^4) 
   \end{equation}  
    Inspection of (7.9) and (7.10) demonstrates the impossiblity of uniform circular motion under the
    action of retarded  Li\'{e}nard -- Wiechert fields. There is a manifest breakdown of Newton's third
    law, that cannot, due to the non-factorisable force components on the right sides of (7.9) and (7.10),
    be restored, as in the discussion of the previous section, by the introduction of different effective
    masses for the interacting objects. Also, since the resultant force is non-central, an unbalanced 
   torque acts on the system. The unbalanced force is:
    \begin{eqnarray}
  \vec{F}_{ret}(1)+\vec{F}_{ret}(2) & = & -\hat{\imath} q^2 \left[\frac{\beta_1+ \beta_2}{r_1+r_2}
     \left(\frac{\beta_2^2}{r_2}
   -\frac{\beta_1^2}{r_1}\right)+\frac{1}{3} \left(\frac{\beta_2^3}{r_2^2}-\frac{\beta_1^3}{r_1^2}\right)\right]
    \nonumber \\
     & + & \frac{\hat{\jmath} q^2(\beta_1^2-\beta_2^2)}{2(r_1+r_2)^2}+O(\beta_1^4,\beta_2^4)
    \end{eqnarray}
     The clockwise (accelerating) torque is of magnitude:
    \begin{equation}
    \Gamma = q^2 \beta_1 \beta_2 \left[\frac{2(\beta_1 +\beta_2)}{r_1+r_2}+\frac{1}{3}\left(\frac{\beta_2}{r_2}+
 \frac{\beta_1}{r_1}\right)\right]+O(\beta_1^6,\beta_2^6)
    \end{equation}
     In ~\cite{Eddington} Eddington discussed the case of the (almost circular) orbits of Jupiter 
     and the Sun under the action of retarded gravitational fields. A simple geometrical argument
      suggests that Jupiter and the Sun would then be subject to  accelerating torques.
    Eddington then considered the analogous electromagnetic problem, and citing the Heaviside
    formula (3.4) for the electric field of a moving charge, stated that the
    direction of the retarded force is `very nearly' in the direction of the centre of mass 
    in this case. It was then concluded that there is no unbalanced torque problem for the 
    Sun-Jupiter system, and that it may be generally assumed that the gravitational force 
    propagates at the speed of light. The calculation presented above shows this conclusion to
    be incorrect.  As the formulae (6.1) and (6.2) follow from only the inverse-square force
    law, SR and Hamilton's Principle, similar formula, with the relacement
     $q_1 q_2 \rightarrow - G m_1 m_2$,
     are expected to hold in Newtonian gravitation, when special relativistic corrections are included.
     In this case also, stable circular orbits cannot be obtained with retarded gravitational fields.
     \par A calculation considering the stability of circular orbits for the case of gravitational
      forces has been previously published~\cite{TVF}. As for the electromagnetic forces just
      considered, and for similar geometrical reasons, an accelerating torque was found in the case
      of retarded gravitational
      fields. By comparing the predicted rate of change of their periods with precision data
     on the rate of change of the observed periods of binary pulsars, a lower limit of 2800c
     for the speed of propagation of gravitational force fields was set. The conclusions
     of Ref.~\cite{TVF} were questioned in Refs.~\cite{Carlip,IPS}. However, the argumentation
     of these authors was based on an analogy between retarded gravitational and electric
     force fields. They concluded that since the retarded LW electric field of
     Eqn(3.4) is radial at the `present time' there is no torque. As discussed above, a similar
     remark was made
     by Eddington~\cite{Eddington}. This conclusion is refuted by the calculation
     presented above. The present-time retarded electric field is radial only for
     a source charge in straight-line motion. As the above calculation and the analogous
     gravitational one of Ref.~\cite{TVF} show, the force is no longer radial when the circular
     geometry of the orbits is properly taken into account. Carlip's argument in Ref.~\cite{Carlip}
     has also been rebutted in Ref.~\cite{TVFV}.

\SECTION{\bf{Summary}}
  The present paper is a sequel to~\cite{JHF1} where inter-charge forces in classical electrodynamics
  were derived from Coulomb's law, SR and Hamilton's Principle. An essential aspect of the derivation
  is the instantaneous nature of the electromagnetic forces, demonstrated in ~\cite{JHF1} to be a prediction
  of QED. In standard text books on classical electromagnetism~\cite{PP,LLCF,Jackson} no clear distinction
    is made between fields describing mechanical forces (the effect, in QED, of the exchange of
   space-like virtual photons) and fields giving a classical description of radiative processes (in QED,
   the creation or destruction of real, on-shell, photons). The importance of such a distinction, the
   former fields being instantaneous, the latter propagating at the
  speed of light, was pointed out 
    many years ago by Chubykalo and Smirnov-Rueda~\cite{CSR}. This conjecture is fully supported by the 
    work presented in~\cite{JHF1} and the present paper. 
   \par The distinction between the standard text-book formulae, denoted in the present paper by the
   labels CEM or LW (for Li\'{e}nard-Wiechert) and those  (denoted by RCED) derived in ~\cite{JHF1} or
   Section 2 above, is most succinctly expressed in terms of the electromagnetic potentials generated 
   by uniformly
  moving electric charges in the different theories. In CEM, no distinction is made between fields
   describing mechanical forces and radiation processes. In both cases the electromagnetic potential
  is assumed to be of the form first derived by Li\'{e}nard and Wiechert~\cite{LW}:
  \begin{eqnarray}
   A^0(CEM) & = & \left\{\frac{q}{(r-\frac{\vec{v} \cdot \vec{r}}{c})}\right\}_{t_{ret}}  \\
  \vec{A}(CEM) & = & \left\{\frac{q \vec{v}}{(r-\frac{\vec{v} \cdot \vec{r}}{c})}\right\}_{t_{ret}}
 \end{eqnarray}
  where all quantities in the large curly brackets are evaluated at the retarded time $t_{ret}= t-r/c$. 
  In RCED there are two physically  distinct types of fields. The first, describing mechanical
   inter-charge forces are derived, using (2.1) and (2.2), from the potentials:
  \begin{eqnarray}
   A^0_{for}(RCED) & = & \frac{q \gamma}{r} = \frac{q u^0}{c r}  \\
  \vec{A}_{for}(RCED) & = &  \frac{q \gamma \vec{v}}{c r} = \frac{q \vec{u}}{c r}
 \end{eqnarray}
  where $u \equiv (\gamma c:\gamma \vec{v})$ is the 4-vector velocity of the moving charge. These potentials
  as well as the associated fields and forces are instantaneous. The retarded potentials of RCED, describing 
   radiation, are the same as (8.3) and (8.4) above, except for a retarded time argument.
 The different physical interpretations of the fields derived from $A_{for}$ and  the retarded
  potential are discussed
  in~\cite{JHF1}. See also~\cite{CSR}.
  It is perhaps surprising, 100 years after the advent of SR, that the formulae (8.1) and (8.2) are still the 
   generally accepted ones for the electromagnetic potential of a moving charge. Indeed $A^0$ and $\vec{A}$
   in these equations are manifestly {\it not} components of a 4-vector, and so it is not to be expected
   that the fields and forces derived from them will respect SR. It has been demonstrated in the present paper
   that indeed they do not, whereas the RCED fields and forces do.
   \par In fact since (8.1) and (8.2) were derived by Li\'{e}nard and Wiechert some seven years before
   Einstein's first SR paper, it would be little short of miraculous if they constituted a 4-vector.
   A similar remark applies to the consistency with SR of the `present time' formula (3.4) for the
   retarded electric field of
   a uniformly moving charge. This may be derived from (8.1) and (8.2), but was first obtained by
    Heaviside~\cite{Heaviside} in 1888. It may be noted, on the other hand, that a velocity-dependent
    scalar potential, consistent with (8.3) up to O($\beta^2$) was proposed as early as 1861 by
    Riemann~\cite{Riemann}\footnote{ The same Bernhard Riemann (1826-1866) is better known for his
  hypothesis concerning the distribution of prime numbers, for pioneering applications of topology
    to the theory of complex functions and, with Lobatchevsky, the invention of
    non-Euclidean geometry. Together with Weber and Clausius, he attempted to construct an
    action-at-a-distance theory of electromagnetic forces in which the fundamental physical objects
   were charged particles, not fields. See the second reference in \cite{Riemann}. This is also the 
   approach adopted in \cite{JHF1} by the present author. Such theories were generally superseded
   in the second half of the 19th century by Maxwell's electromagnetic field theory.}. 
     \par The mechanical forces in CEM derived from (8.1) and (8.2) have been compared above with those of
    RCED derived from (8.3) and (8.4). In Section 3 it is shown that use of the CEM formula (3.4) to calculate
    small-angle Rutherford scattering gives a scattering angle smaller by a factor $1/\gamma$ than the
    well-known standard result, that is recovered by use of the RCED formula (3.3). It is also shown in Section 3
    that that conservation of relativistic transverse momentum requires that transverse forces in 
    different inertial frames scale according to $1/\gamma$ (see Eqn(3.17)).
    This condition is respected by the RCED fields but not
     by the CEM ones. In Section 4 it is shown that the flux of the electric field of a moving charge in
    RCED does not (unlike in CEM where the electric flux is the same for a charge at rest or in motion)
     respect Gauss' Law. The flux depends both on the velocity of the charge and the geometry of
     the bounding surface. The electric field is a mathematical abstraction useful to simplify
     dynamical formulae expressed in terms of forces, but has, unlike charges and their motion,
     no objective existence. In Section 5, Jackson's `torque paradox'~\cite{JackTP} for two interacting charges,
     where electric and magnetic fields are derived from the CEM potentials (8.1) and (8.2)
     is discussed. A non-vanishing torque is predicted in one inertial frame, a vanishing one
     in another, in contradiction with the special relativity principle. It is shown that use
     of RCED fields gives a vanishing torque in all inertial frames, consistent with SR.
     In Section 6 it is demonstrated that stable, circular, Keplerian orbits are possible 
     for two oppositely-charged objects using exact RCED dynamical formulae. An exact
     relativistic generalisation of Kepler's third law relating the orbit period to
     the separation of the charges is derived for this case. In Section 7 the impossiblity
      of stable circular motion under forces derived from retarded CEM fields is demonstrated.
      Such forces do not respect Newton's third law and, as already conjectured during the
      19th Century for the analogous gravitational case~\cite{Eddington} and recently
      demonstrated by explicit calculation~\cite{TVF}, generate an unbalanced
     torque. Stable orbital motions of electrons in atoms or of planets around the Sun are only 
      possible if (as Newton assumed for the case of gravity) the inverse-square-law electrical        
      or gravitational forces act instantaneously.

\newpage
               
 {\bf Appendix}
 \par  Denoting by $\delta t$ the time of passage of O$_2$ from B to E (Fig.5) then:
  \[\delta t = \frac{r}{c} = \frac{\phi r_2}{v_2}~~~~~~~~~~~~~~~~~~~~~~~~~~~~~(A1)\]
   From the geometry of Fig.5:
     \begin{eqnarray}
     r & = & r_1\cos \phi_1 + r_2 \cos \phi_2~~~~~~~~~~~~~~~~~~~~~~~(A2) \nonumber \\
     r_1 \sin \phi_1  & = &  r_2 \sin \phi_2~~~~~~~~~~~~~~~~~~~~~~~~~~~~~~~~~~~~~(A3) \nonumber \\
      \phi & = & \phi_1 + \phi_2 ~~~~~~~~~~~~~~~~~~~~~~~~~~~~~~~~~~~~~~~(A4) \nonumber \\
      r_1 v_2 & = &  r_2 v_1 ~~~~~~~~~~~~~~~~~~~~~~~~~~~~~~~~~~~~~~~~~~~(A5) \nonumber 
    \end{eqnarray}
    It follows from (A1-A5) that: 
    $\phi = \beta_1 + \beta_2$, $\phi_1 = \beta_2$+O($\phi_1^3$) and $\phi_2 = \beta_1$+O($\phi_2^3$) .  
  \par  Making the small angle approximations:
   \[ \sin \phi \simeq \phi-\frac{\phi^3}{6}~~~~\cos \phi \simeq =1 -\frac{\phi^2}{2} \]
 equations (A2)-(A5) may be combined to give:  
     \begin{eqnarray}
     r & \simeq & (r_1+r_2)(1-\frac{r_1 \beta_2^2}{2 r_2})  \nonumber \\
       & = &  (r_1+r_2)(1-\frac{\beta_1\beta_2}{2})+ O(\beta_2^3)~~~~~~~~~(A6) \nonumber \\
     r-\vec{\beta_2} \cdot \vec{r}  & = & r(1+\beta_2 \sin \phi_2) \nonumber \\
      & \simeq &  r(1+\beta_1 \beta_2) \nonumber \\
       & = & (r_1+r_2)(1+\frac{\beta_1\beta_2}{2})+ O(\beta_2^3)~~~~~~~~~(A7) \nonumber 
    \end{eqnarray}
   also 
   
  \[ \vec{r}' \equiv \vec{r}-\vec{\beta_2}r = r'(\hat{\imath} \sin \alpha-\hat{\jmath} \cos \alpha) 
    ~~~~~~~~~~~~~~~(A8)\]
    Substituting (A6)-(A8) into (7.3) and noting that $q_2 = -q$ and $r' = r \sec(\phi_1 - \alpha)$
    gives:  
     \begin{eqnarray} 
      \vec{E}_{ret}^{vel}(1) & \simeq & \frac{q(1-\beta_2^2)(-\hat{\imath} \sin \alpha+\hat{\jmath} \cos \alpha)}
         {r^2(1+\beta_1\beta_2)^3 \cos(\phi_1 - \alpha)} \nonumber \\
     & \simeq &  \frac{q(1-\beta_2^2)(1+\frac{\beta_2^2}{2})(-\hat{\imath} \sin \alpha+\hat{\jmath} \cos \alpha)}
         {(r_1+ r_2)^2(1+2\beta_1\beta_2)} \nonumber \\
     &  =  &   \frac{q[-\hat{\imath} \alpha+\hat{\jmath}\{1-\beta_2(2 \beta_1+\beta_2/2)\}]}
           {(r_1+ r_2)^2}+O(\beta_1^4,\beta_2^4)~~~~~~~~~(A9) \nonumber 
     \end{eqnarray} 
     The last member follows since, as shown below, $\alpha$ is of order $\beta_2^3$.
  \newpage
    From the geometry of Fig.5:
      \begin{eqnarray} 
     \tan \alpha & = & \frac{CD \cos \phi}{r_1+r_2\sec \phi-CD\sin \phi}  \nonumber \\      
     & = & \frac{(r_2 \tan \phi - r \beta_2)\cos \phi}{r_1+r_2\sec \phi- (r_2 \tan \phi - r \beta_2)\sin \phi}
      \nonumber \\
    & = & \frac{r_2( \sin \phi - \phi  \cos \phi)}{r_1+r_2\sec \phi- r_2(\tan \phi - \phi)\sin \phi}
    ~~~~~~~~~~~~(A10)  \nonumber 
     \end{eqnarray}
 where (A1) has been used. 
   Noting that 
   \[  \sin \phi - \phi  \cos \phi = \phi-\frac{\phi^3}{6}-\phi + \frac{\phi^3}{2}+...
   = \frac{\phi^3}{6}+O(\phi^5) \]
 \[  \tan \phi \sin \phi - \phi  \sin \phi = \frac{(\phi-\frac{\phi^3}{6})^2}{1-\frac{\phi^2}{2}}
   -\phi^2+\frac{\phi^4}{6}+...=\frac{\phi^4}{3}+ O(\phi^6) \]
 \[ \sec \phi = 1+\frac{\phi^2}{2}+... \]
 (A10) gives the result:
  \[ \alpha \simeq \arctan \left[\left(\frac{r_2}{r_1+r_2}\right) \frac{\phi^3}{3}\right] =
   \left(\frac{r_1+r_2}{r_2}\right)^2 \frac{\beta_2^3}{3}+O(\beta_2^5)~~~~~~(A11) \]
    Substitution of $\alpha$ given by (A11) into (A9) gives Eqn(7.6) of the text.
  \par Since the modulus of the velocity vector of an object in uniform circular
    motion is constant, the increment $\delta \vec{\beta_2}$ for O$_2$ is  anti-parallel to 
    the radius vector $\vec{r}_2$. It then follows from Fig.5 and the geometrical conditions (6.6) that:
     \begin{eqnarray} 
  \vec{r} \times [(\vec{r}-\vec{\beta_2}r)
  \times d\vec{\beta_2}/dt] &  =  & \frac{r r' \beta_2 v_2 \sin \phi }{r_2}[ \hat{\imath} \cos \phi_1+
   \hat{\jmath} \sin \phi_1] + O(\beta_2^4)~~~~~~~~~~~~~~(A12) \nonumber
     \end{eqnarray}
     Subsituting (A12) and (A7) into (7.4) gives:  
      \begin{eqnarray}
  \vec{E}_{ret}^{acc}(1) & \simeq & -\frac{q \beta_2^2 (\beta_1+\beta_2) [ \hat{\imath} \cos \phi_1+ \hat{\jmath} \sin \phi_1]}
    {r r_2 (1+\beta_1\beta_2)^3 \cos(\phi_1 - \alpha)} \nonumber \\ 
 & \simeq & -\frac{q \beta_2^2 (1+\frac{\beta_2^2}{2})(\beta_1+\beta_2)
  [ \hat{\imath} (1-\frac{\beta_2^2}{2})+ \hat{\jmath}
 \beta_2 ]}
    {(r_1+r_2)r_2 (1+\beta_1\beta_2)^3(1-\frac{\beta_1\beta_2}{2})}
 \nonumber~~~~~~~~(A13)
     \end{eqnarray}
  which yields Eqn(7.7) of the text on retaining only terms up to $\beta_2^3$, $\beta_2^2 \beta_1$.     
\pagebreak


\begin{thebibliography}{99}
\bibitem{Feyn1} 
R.P.Feynman, R.B.Leighton and M.Sands, `The Feynman Lectures in Physics'
 (Addison-Wesley, Reading Massachusetts, 1963), Vol 1 Ch 28-1, `Electromagnetism I' Ch 12-7.
\bibitem{QED} See for example the reviews in: T.Kinoshita, `Quantum Elecrodynamics',
   (World Scientific, Singapore 1990).
 \bibitem{Feyn2}
 R.P.Feynman, `QED The Strange Theory of Light and Matter', 
  (Princeton University Press, 1985) Chapter 3.
 \bibitem{JHF1}
J.H.Field, Phys. Scr. {\bf 74} 702 (2006),
 http://xxx.lanl.gov/abs/physics/0501130.
 \bibitem{Kohletal}
A.L.Kholmetskii {\it et al.} J. Appl. Phys. {\bf 101} 023532 (2007).
 http://xxx.lanl.gov/abs/physics/0601084.
\bibitem{JHFnrf}
 J.H.Field, `Quantum electrodynamics and experiment demonstrate
   the non-retarded nature of electrodynamical force fields',
   http://xxx.lanl.gov/abs/0706.1661.
  \bibitem{Hertz}
H.Hertz, Ann. der Physik {\bf XXXIV} 551 (1888).  
 \bibitem{PP}
 W.H.Panofsky and M.Phillips, `Classical Electricity and Magnetism', 2nd Edition
  (Addison-Wesley, Cambridge Mass, 1962) Ch 19.
 \bibitem{LLCF}  
 L.D.Landau and E.M.Lifshitz `Classical Theory of Fields', 
  Translated by M.Hamermesh, (Pergamon Press, Oxford, 1962).
  \bibitem{Jackson}
 J.D.Jackson, `Classical Electrodynamics',
 Second Edition (John Wiley and Sons, New York, 1975).
 \bibitem{RHR}
 R.H.Romer, Am. J. Phys. {\bf 34} 772 (1966).
 \bibitem{SJ}
 W.Shockley and R.P.James, Phys. Rev. Lett.  {\bf 18} 876 (1967). 
 \bibitem{CVV}
 S.Coleman and J.H. Van Vleck, Phys. Rev. {\bf 171} 1370 (1968).
 \bibitem{Furry}
 W.H.Furry, Am. J. Phys. {\bf 37} 621 (1969).
 \bibitem{GGL}
 G.G.Lombardi, Am. J. Phys. {\bf 51} 213 (1983).
 \bibitem{VH}
 V.Hnizdo, Am. J. Phys. {\bf 51} 213 (1983).
\bibitem{ODJ}
 O.D.Jefimenko, Eur. J. Phys. {\bf 20} 39 (1999).\
\bibitem{CE}
 See, for example, B.I.Bleaney and B.Bleaney `Electricity and Magnetism' (O.U.P., Oxford, 1957) Chapter XIX,
 or C.Kittel, `Introduction to Solid State Physics', 6th Edition (J.Wiley and Sons New York 1986) Chapter 7.
 \bibitem{Eddington}
A.S.Eddington, `Space, Time and Gravitation, An Outline of
  General Relativity', (CUP, Cambridge, 1920) Chapter VI.
  \bibitem{EinsteinGR}
 A.Einstein, Ann. Phys. (Leipzig)  {\bf 49} 767 (1916).
 \bibitem{JackRS}
Reference~\cite{Jackson} above, Chapter 13, P619. 
\bibitem{LW}
 A.Li\'{e}nard, L'Eclairage Electrique, {\bf16} pp5, 53, 106 (1898);
  \newline E.Wiechert, Archives N\'{e}land (2)  {\bf5} 459 (1900).
\bibitem{Heaviside}
O.Heaviside, The Electrician,  {\bf 22} 1477 (1888),
 Philos. Mag. {\bf 27} 2324 (1889). 
 \bibitem{PDGIL}
Review by D.E.Groom and S.R.Klein in: `Review of Particle Properties', 
 S.Eidelman {\it et al}, Phys. Lett. {\bf 592} 242 (2004).
 \bibitem{JackLC}
Reference~\cite{Jackson} above, Section 11.10.
 \bibitem{JackTP}
J.D.Jackson, Am. J. Phys. {\bf 72} 1484 (2004).
 \bibitem{PPTN}
Ref.~\cite{PP} above. Section 15.2, P274.
 \bibitem{TN}
F.T.Trouton and H.R.Noble, Phil. Trans. {bf A202} 165 (1903),
 Proc. Roy. Soc. {\bf 72} 132 (1903).
 \bibitem{LLCFLW}
 Reference~\cite{LLCF}, Chapter 8, Section 63, Equations (63.8-9).
\bibitem{TVF}
 T.Van Flandern, Phys. Lett. {\bf A250} 1 (1998).
\bibitem{Carlip}
 S.Carlip, Phys. Lett. {\bf A267} 81 (2000).
\bibitem{IPS}
 M.Ibison, H.E.Puthoff and S.R.Little, `The speed of gravity revisited'
     http://xxx.lanl.gov/abs/physics/9910050.
\bibitem{TVFV}
 T.Van Flandern and J.P.Vigier, Foundations of Physics {\bf 32} 1031 (2002).
\bibitem{CSR}
 A.E.Chubykalo and R.Smirnov-Rueda, Phys. Rev. {\bf E53} 5373 (1996).
\bibitem{Riemann}
B.Riemann, Lectures `Schwere, Elekricitat und Magnetismus', Hannover, 1875, P326.
See also E.T.Whittaker, `A History of Theories of Aether and Electricity',
 (Am. Inst. Phys., New York, 1987) Chapter 7.
\end{thebibliography}
\end{document}